\documentclass{article}[12pt]

\usepackage{epsfig,amsfonts,amssymb,newlfont,rotate,float,array}

\newcommand{\bbox}[1]{\mbox{\boldmath $#1$}} 
\newcommand{\sbbox}[1]{\mbox{\boldmath \scriptsize $#1$}} 

\begin{document}

\title{Scaling corrections: site-percolation
and Ising model\\ in three dimensions}

\author{
H.~G.~Ballesteros$^{\mathrm a}$\footnote{\tt hector@lattice.fis.ucm.es},
L.~A.~Fern\'andez$^{\mathrm a}$\footnote{\tt laf@lattice.fis.ucm.es},
V.~Mart\'{\i}n-Mayor$^{\mathrm a}$\footnote{\tt victor@lattice.fis.ucm.es},\\
A.~Mu\~noz Sudupe$^{\mathrm a}$\footnote{\tt sudupe@lattice.fis.ucm.es},
G.~Parisi$^{\mathrm b}$\footnote{\tt giorgio.parisi@roma1.infn.it}, and
J.~J.~Ruiz-Lorenzo$^{\mathrm a}$\footnote{\tt ruiz@lattice.fis.ucm.es}.\\
$^{\mathrm a}${\it Departamento de F\'{\i}sica Te\'orica, 
Universidad Complutense de Madrid},\\
{\it  28040 Madrid, Spain,} \\
$^{\mathrm b}${\it Dipartimento di Fisica, Universit\`a di Roma I},\\ 
{\it P. A. Moro 2, 00185 Roma, Italy,}
}
\date{May 14, 1998}
\maketitle

\begin{abstract}
Using Finite-Size Scaling techniques we obtain accurate results for
critical quantities of the Ising model and the site percolation, in
three dimensions. We pay special attention in parameterizing the
corrections-to-scaling, what is necessary to put the systematic errors
below the statistical ones.
\end{abstract}

\noindent {\it PACS:}
75.40 Mg,   
75.50.Lk.   
05.50.+q,   
75.40.Cx.   

\section{Introduction}

The concept of Universality is perhaps one of the main discoveries of
Modern Physics~\cite{UNIVERSALIDAD}.  The critical exponents of phase
transitions are among the most important quantities in Nature, as they
offer the most direct test of Universality.  Therefore, precise
experimental measures of these exponents combined with accurate
theoretical calculations are crucial cross-checks. Unfortunately, in
three dimensions the range of variation of the exponents is very
narrow. For instance, the correlation-length exponent, $\nu$, varies
within a 10\% interval for most 
systems~\cite{ZINN-JUSTINLIBRO}.
Therefore, in order to distinguish between different universality
classes, it is necessary to measure or calculate these quantities with
several significant figures.

There exist some powerful analytical techniques for computing
critical exponents:
$\epsilon$-expansions, high-$T$ series, $N$-expansions, or perturbative
expansions at fixed dimension. A recent and complete study on this kind of
calculations can be found in Ref.~\cite{ZINN-JUSTIN}. A drawback of
this approach is that the error estimate is quite involved.
However, a 0.15\% precision can be reached for $\nu$ in Ising systems.

A competing alternative is the use of Finite-Size Scaling (FSS)
techniques~\cite{BARBER}  combined with a Monte Carlo (MC) method,
which in principle is able to measure with unlimited precision. The
FSS method has the remarkable property of using the finite size
effects to extract information about the critical properties of the
system. In the language of the Renormalization Group (RG), we expect
that for large enough lattices the divergences are fully described by
the relevant operators.  The MC method itself is not quite efficient as
the statistical errors in measures decrease only as the inverse square
root of the numerical effort.  However, the present sophisticated
numerical techniques and algorithms, as well as the high computer
power available, have allowed to reduce the statistical fluctuations
largely.  One could naively think that to get one more significant
digit for a critical exponent is only a matter of multiplying by 100
the CPU time.  This is not true in general, since the effects due to
the finite-size of the simulated lattices eventually become larger
than the statistical errors (in the RG language, the effects induced by the
irrelevant couplings cannot be neglected anymore).  Traditionally, one
designs a simulation in order to get the systematic errors to lie below
the statistical ones. With very high precision, a more quantitative
treatment of systematic errors is required. 

In this work, we want to
deal with the leading irrelevant terms (or corrections-to-scaling
terms in the FSS language) in two of the simplest models in three
dimensions: the Ising model and the
site percolation~\cite{STAUFFERBOOK}. The reader might be surprised
that the FSS ansatz holds for such a simple model as percolation,
which is essentially not dynamical. The underlying reason is that
bond-percolation is the $q\to 1$ limit of the $q$-states Potts model,
as can be seen through the ``Fortuin-Kasteleyn'' representation of the
latter~\cite{KAFO}.  The importance of both models has justified the
construction of specific hardware as the Ising computer at Santa
Barbara~\cite{ISINGCOMP}, Percola~\cite{PERCOLA} or
the Cluster Processor~\cite{TALAPOV}. However, the present update
methods, as well as the power of the computers available allow to
obtain very accurate measures for Ising models in general purpose
computers. Regarding the percolation, an useful technical development
has been the introduction of a reweighting method~\cite{PERC,HARRIS},
which allows to extrapolate the simulation results obtained at
dilution $p$ 
to a nearby $p'$ dilution. As an outcome, dilution-derivatives can be
also efficiently measured. This has suggested a different simulation
strategy from the usual in percolation
investigations~\cite{PERCOLA,LORENZ,GRASSBERGER,STAUFFER}.  Instead of
producing a small number of very large samples, we generate ${\cal O}
(10^7)$ different samples in smaller lattices, in order to accurately
measure derivatives with respect to the dilution and to obtain accurate
extrapolations. The very nice agreement~\cite{PERC} with supposedly
exact results for the critical exponents in two
dimensions~\cite{NIENHUIS}, and with other numerical results in three
dimensions (see table ~\ref{OTROSPERC}), allows a great confidence in this new
approach. In addition, the coincidence of two algorithmically
different studies is a cross-check that reinforces {\it both}.

The specific FSS method we use in this paper is based on comparison of
measures taken in two different lattices at the value of the
``temperature'' for which the correlation-length in units of the
lattice size is the same for both~\cite{OURFSS,PERC}. 
Comparatively, this method is
particularly well suited for the measure of magnetic critical
exponents and for the parameterization of the effects induced by
the irrelevant operators.  We shall show that at the precision level
we can reach (as small as 0.1\% for the critical exponent $\nu$,
extracted from a given lattice pair), to take into account the effect
of the leading irrelevant operator is unavoidable.  For the two simple
models we consider, very different situations are found. For
site percolation, the scaling corrections exponent, $\omega$, is so
large ($\omega\approx 1.6$) that other commonly ignored corrections,
such as the induced by the analytic part of the free-energy, are of
the same order. This makes our estimates of the critical exponents
quite independent of the details of the infinite-volume extrapolation.
But, on the other hand, the
parameterization of the scaling corrections is remarkably difficult.
On the contrary, for the Ising model we have $\omega\approx 0.8$, and
the infinite-volume extrapolation is mandatory.  
But the critical exponents related with higher-order
corrections are large enough to allow for a neat, simple
parameterization.

In the next section we shall describe the FSS method we use. 
The measured observables are defined in section 3.
The results for the
Ising model and the site percolation are reported 
in sections 4 and 5, respectively. We will finish with the conclusions.

\section{Finite-Size Scaling}

Nowadays, a nice unifying picture of critical phenomena is provided
by the  Renormalization Group. In this frame, one can study not only
the leading singularities defining the critical exponents, but also
subdominant corrections (the Wegner confluent corrections~\cite{WEGNER}).
In addition, from the Renormalization Group, a transparent 
derivation of the Finite-Size Scaling Ansatz (FSSA)  follows (see~\cite{BARBER}
and references therein).
The starting point is the 
free energy of  a $d$-dimensional system
\begin{equation}
f(t,h,\{u_j\})=g(t,h,\{u_j\})
+b^{-d} f_{\mathrm {sing}}(b^{y_t} t, b^{y_h} h, 
\{u_j b^{y_j} \}),
\label{rg_whole}
\end{equation}
where $f_{\mathrm {sing}}$ is the so-called singular part, while $g$ is an 
analytical function. We call $b$ to the block size in the Renormalization Group
Transformation (RGT), while $y_t$, $y_h$ and $y_j$ 
($j \ge 3$) are the
eigenvalues of the RGT with scaling fields $t$, $h$ and $u_j$ ($j \ge
3$). In the simplest applications (such as the ones we are
considering) there are two relevant parameters: the
``thermal field'', $t$, and the magnetic field, $h$ ({\it i.e. } 
$y_t>0, y_h>0$) and
we denote by $\{u_j\}$  the set of the irrelevant
operators ($0\ge y_3 \ge y_4 \ge y_5 \ge\ldots$).
One commonly uses the definitions $\nu=1/y_1$, $\eta=2+d-2y_h$ and 
$\omega=-y_3$. The
scaling field $t$ can be identified with the reduced temperature in
Ising systems, or with $(p-p_{\mathrm c})/p_{\mathrm c}$ in
percolation problems.
Taking derivatives of the free
energy with respect to $t$ or
$h$ it is possible to compute the critical behaviour of the different 
observables, including their scaling corrections~\cite{WEGNER}. A very
similar strategy is followed in the study of a finite lattice, where
we write for the free energy 
(see~\cite{BLOTE} for a detailed presentation)
\begin{equation}
f(t,h,\{u_j\}, L^{-1})=g(t,h,\{u_j\})
+b^{-d} f_{\mathrm {sing}}(b^{y_t} t, b^{y_h} h, 
\{u_j b^{y_j} \}, b/L).
\label{rg_L}
\end{equation}
At this point one takes $b=L$, thus arriving to a single-site lattice.
By performing the appropriate derivatives, all the critical quantities
can be computed. The result can be cast in general form for a
quantity $O$ diverging like $t^{-x_O}$ in the thermodynamical limit:
\begin{equation}
O(L,t)=L^{x_O/\nu}\left[F_O\left(\frac{L}{\xi(\infty,t)}\right) +
{\cal O}(L^{-\omega},\xi^{-\omega})\right]\, ,
\label{FSS1}
\end{equation}
where $F_O$ is a smooth scaling function. In usual applications 
one is interested in the $\xi\gg L$ regime,
thus $\xi^{-\omega}$ is safely neglected. Of course in
Eq.~(\ref{FSS1}), we have only kept the leading irrelevant eigenvalue,
but,
in fact, other scaling corrections like
\begin{equation}
\{L^{y_j}\},\{L^{y_j+y_i}\},\ldots\  (i,j \ge 3)
\label{HCORR}
\end{equation}
are to be expected.  In addition, other kind of terms are induced by
the analytical part of the free energy, $g$. For the susceptibility
(or related quantities like the Binder cumulant or the
correlation-length, see below) one should take the second derivative
with respect to the magnetic field, $h$, in Eq.~(\ref{rg_L}). The
leading contribution of the analytical part is independent of the
lattice size, thus if one wants to cast the result as in
Eq.~(\ref{FSS1}), corrections like $L^{-\gamma/\nu}$ should be added.

Equation~(\ref{FSS1}) is still not convenient for a numerical
study, because it contains not directly measurable quantities like 
$\xi(\infty,t)$. Fortunately, it can be turned into an useful expression 
if a reasonable definition of the correlation length in a finite
lattice, $\xi(L,t)$, is available:
\begin{equation}
O(L,t)=L^{x_O/\nu}\left[\widetilde F_O\left(\frac{\xi(L,t)}{L}\right)
+ {\cal O}(L^{-\omega})\right]\ ,
\label{FSS}
\end{equation}
where $\widetilde F_O$ is a smooth function related with $F_O$ and
$F_\xi$.

To reduce the effect of the corrections-to-scaling terms, one could
take measures only in large enough lattices. Even in the simplest
models, as the two considered in this paper, if one wants to obtain
very precise results, the lattice sizes required can be
unreachable. However, we shall show that this is not the most
efficient option.
In the specific method we use, the scaling function is 
eliminated by taking measures of a given observable at the same
temperature in two different lattice sizes ($L_1,L_2$). At the
temperature where the correlation lengths are in the ratio $L_1:L_2$,
from Eq.~(\ref{FSS}) we can write the quotient of the measures 
of an observable, $O$, in
both lattices as
\begin{equation}
\left.Q_O\right|_{Q_\xi=\frac{L_1}{L_2}}=\left(\frac{L_1}{L_2}\right)^{x_O/\nu}
+A_{Q_O} L_2^{-\omega}+\ldots\ ,
\label{QUO}
\end{equation}
where $A_{Q_O}$ is a constant.

The great advantage of Eq.~(\ref{QUO}) is that to obtain the
temperature where ${Q_\xi=L_1/L_2}$, only two lattices are required,
and a very accurate and statistically clean measure of that
temperature can be taken. In addition, the statistical correlation 
between $Q_O$ and $Q_\xi$ reduces the fluctuations. Other methods, such as
measuring at the peak of some observable suffer in general from larger
corrections-to-scaling. Computing the infinite volume critical
temperature and measuring at that point performs well for studying
observables that vary slowly at the critical point, as those used
for computing the $\nu$ exponent. However, the magnetic exponents
require measuring quantities that change rapidly with the temperature
and this is more involved. We think that our method outperforms any
other previously used, specially in the computation of the $\eta$
exponent. 

To perform an extrapolation following Eq.~(\ref{QUO}), an estimate 
of $\omega$ is required. This can
be obtained from the behaviour of dimensionless quantities, like the
Binder cumulant or the correlation length in units of the lattice
size, $\xi(L,t)/L$, which remain bounded at the critical point although
their $t$-derivatives diverge. For a generic
dimensionless quantity, $g$, we shall have a crossing 
$$g(L, t^{\mathrm {cross}}(L,s))=g(sL, t^{\mathrm {cross}}(L,s)).$$
The distance from the critical point, $t^{\mathrm {cross}}(L,s)$,
goes to zero as~\cite{BINDER}:
\begin{equation}
t^{\mathrm {cross}}(L,s)
\propto \frac{1-s^{-\omega}}{s^{1/\nu}-1}L^{-\omega-1/\nu}\ .
\label{BETACFIT}
\end{equation}
From Eq.~(\ref{BETACFIT}), a clean estimate of $\omega$ can be
obtained provided that $|y_4|-\omega$ and $\gamma/\nu - \omega$ are
large enough (say of order one).

\section{The Models}

We will consider a cubic lattice with periodic boundary conditions
and linear size $L$, the volume being $V=L^3$. In
the case of the Ising model we consider the usual Hamiltonian 
\begin{equation}
H=-\beta \sum_{<i,j>}\sigma_i \sigma_j\ ,
\label{ISINGHAMIL}
\end{equation}
where the sum is extended over nearest neighbour sites and the spin
variables are $\pm 1$.

The fundamental observables we measure are the energy, and the magnetization
\begin{equation}
 E =\sum_{<i,j>}\sigma_i\sigma_j\ , \qquad {\cal
M}=\frac{1}{V}\sum_i \sigma_i\ .
\label{ENERMAG}
\end{equation}

The energy is extensively used for $\beta$
extrapolation \cite{REWEIGHT} and for calculating $\beta$-derivatives
through its connected correlation. 

The other quantities that we measure are related with the magnetization.
In practice we are interested in mean values of even powers of the
magnetization as the susceptibility
\begin{equation}
\chi=V{\left\langle {\cal M}^2 \right\rangle}\ ,
\end{equation}
or the Binder parameter
\begin{equation}
g_4=\frac{3}{2}-\frac{1}{2}\frac{{\langle {\cal M}^4\rangle}}
           {{\langle {\cal M}^2 \rangle^2}}\ .
\label{G4}
\end{equation}

The cumulant $g_4$ tends to a finite and universal value at the
critical point.
As correlation-length in a finite lattice, we use a quantity that
only involves second powers of the magnetization, but uses the Fourier
transform of the spin field
\begin{equation}
\widehat\sigma(\bbox{k})=\frac{1}{V}\sum_{\sbbox{r}}{\mathrm e}^{\mathrm i
\sbbox{k}\cdot\sbbox{r}}\sigma_{\sbbox{r}}\ .
\end{equation}
Defining 
\begin{equation}
F=\frac{V}{3}{\left\langle |\widehat\sigma(2\pi/L,0,0)|^2
+\mathrm{permutations}\right\rangle}\ ,
\label{F}
\end{equation}
we will use as correlation length~\cite{COOPER} 
\begin{equation}
\xi=\left(\frac{\chi/F-1}{4\sin^2(\pi/L)}\right)^\frac{1}{2}.
\label{XI}
\end{equation}

The site percolation is defined by filling the nodes of a lattice with
probability $p$. Once the lattice sites are filled (we call this
particular choice a {\it sample}) a system of spins is placed in the
occupied nodes.  The spins interact with the Hamiltonian
(\ref{ISINGHAMIL}) at zero temperature ($\beta=\infty$).  In this way
neighbouring spins should have the same sign, while the signs of spins
belonging to different clusters ({\it i.e.} not connected through an
occupied lattice path) are statistically uncorrelated. Thus, by
counting the number of spins contained in each cluster, $\{n_c\}$, we
know the exact values of $\langle \cal M^2\rangle$ and $\langle \cal
M^4\rangle$ in a particular sample:
\begin{eqnarray}
\langle {\cal M}^2 \rangle &=&\frac{1}{V^2}\sum_c n_c^2,\\\nonumber
\langle  {\cal M}^4 \rangle&=&3\langle {\cal M}^2 \rangle^2
-\frac{2}{V^4}\sum_c n_c^4, 
\label{M2M4}
\end{eqnarray}
where the sums are extended to all the clusters.

To compute the quantities involving Fourier transforms of the magnetization 
we measure
\begin{equation}
\widehat n_c (\bbox{k})=\frac{1}{V}\sum_{\sbbox{r}\in c }{\mathrm e}^{\mathrm i
\sbbox{k}\cdot\sbbox{r}}\sigma_{\sbbox{r}}\ ,
\end{equation}
where the sum is extended to the sites of the $c$-th cluster, arriving
to
\begin{equation}
\langle | \widehat\sigma(\bbox{k}) |^2 \rangle = 
\sum_c  | \widehat n_c (\bbox{k}) |^2 .
\label{FOURT}
\end{equation}
We then average Eqs. (\ref{M2M4}) and (\ref{FOURT}) in the
different samples generated. This new average will be denoted by an
overline. So we define the correlation length and the cumulant $g_4$ as
\begin{eqnarray}
\xi&=&\left(\frac{\overline{\chi}/\overline{F}-1}{4\sin^2(\pi/L)}\right)^\frac{1}{2}, \\
g_4&=&\frac{3}{2}-\frac{1}{2}\frac{{\overline{\langle {\cal M}^4\rangle}}}
           {\overline{{\langle {\cal M}^2 \rangle}}^2}\ .
\end{eqnarray}
Another universal quantity, whose non-vanishing
value proofs that the susceptibility is not a self-averaging quantity,
is the cumulant $g_2$ 
\begin{equation}
g_2=\frac{\overline{ \langle {\cal M}^2 \rangle^2 - 
\overline{\langle {\cal M}^2 \rangle}^2
}}{\overline{ \langle {\cal M}^2 \rangle}^2 } \ .
\end{equation}

A last technical comment for our percolation study is that we store
the actual density values obtained with  probability $p$, in order
to perform a $p$-extrapolation of the mean values of the interesting
observables, and also $p$-derivatives\cite{PERC,HARRIS,ISDIL}.

Both for the Ising model and site percolation, 
the observables we use to compute the two independent
critical exponents, $\eta$ and $\nu$, are 
\begin{eqnarray}
\chi&\rightarrow& x=\nu(2-\eta), \label{OBS} \\ 
\partial_\beta\xi,\, \partial_p\xi &\rightarrow& x=\nu+1\ .\nonumber
\end{eqnarray}

For the sake of completeness we will link here our method
with the more classic approach followed in percolation.
The basic entity in percolation is the number of clusters of size
$s$ divided by
the lattice volume, $n_s$~\cite{STAUFFERBOOK}. This object induces a probability of
finding a cluster of size $s$, given by $s n_s$. Near the 
percolation threshold, $p_{\mathrm c}$, $n_s$ follows the law
\begin{equation}
n_s = s ^{-\tau} f(s^{1/\sigma}(p-p_{\mathrm c})),
\end{equation}
where $\sigma$ and $\tau$ are critical exponents, 
and $f$ is a scaling function. This yields
just at $p=p_{\mathrm c}$
\begin{equation}
n_s = s ^{-\tau}\, (A + B s^{-\Omega}+\ldots)\, ,
\label{NS}
\end{equation}
where $\Omega$ is a corrections-to-scaling exponent.
We can relate the thermodynamical critical exponents,
$\eta$, $\nu$ and $\omega$ with the more standard exponents in percolation
$\sigma$, $\tau$ and $\Omega$: 
\begin{eqnarray}
\nu &=& \frac{\tau-1}{\sigma d} , \\
\eta & =& \frac{(2-d) \tau + 3d -2}{\tau-1}  , \\
\omega& =& \frac {\Omega}{\sigma\nu}\,  , 
\end{eqnarray}
where $d$ is the spatial dimension of the lattice (in opposition
to the fractal dimension $d_{\mathrm f}=(\sigma\nu)^{-1}$).

\section{Results for the Ising model}

\begin{table}[b!]
\begin{center}
\begin{tabular*}{\linewidth}{@{\extracolsep{\fill}}rlll}\hline\hline
\multicolumn{1}{c}{$L$} 
& \multicolumn{1}{c}{$\nu$} 
& \multicolumn{1}{c}{$\eta$} 
& \multicolumn{1}{c}{$g_4$} \\ \hline
8        &   0.64379(37)  &   0.01097(40)   &0.72177(19)\\
12       &   0.63778(46)  &   0.02094(31)   &0.71460(19)\\
16       &   0.63654(40)  &   0.02548(38)   &0.71048(18)\\
24       &   0.63385(44)  &   0.02927(34)   &0.70668(21)\\
32       &   0.63277(48)  &   0.03129(38)   &0.70481(25)\\
48       &   0.63164(48)  &   0.03273(37)   &0.70317(25)\\
64       &   0.6316(6)  &   0.03376(39)   &0.70204(33)\\\hline
$\infty$ &   0.6294(5)(5) &   0.0374(6)(6)  &0.6984(5)(6)\\\hline\hline
\end{tabular*}
\caption{Critical exponents $\nu$ and $\eta$ 
for the Ising model obtained from pairs
of lattices of sizes $(L,2L)$ where $Q_{\xi}=2$. We present also
$g_4(L)$ at the same points. The last row corresponds to an infinite
volume extrapolation considering the leading scaling corrections.
The second error bar is induced by the uncertainty in $\omega$: when
$\omega$ increases, $\nu$ and $g_4$ increase, while $\eta$ decreases.}
\label{TEXPOISING}
\end{center}
\end{table}

We have used a Single Cluster (SC) update algorithm~\cite{WOLFF} which is
known to perform very well for this model. We take measures every 50
SC. We have accumulated 8 millions of measures for lattice sizes
$L$=8, 12, 16, 24, 32, 48, 64, 96  and 128 at $\beta=0.22165$. For the
statistical analysis, we use a jack-knife method with 50 bins of data.
Our pseudo random number generator has been
a corrected shift register generator introduced in
Ref.~\cite{PARISIRAPUANO} improved by adding (modulus 1) a congruential
generator (see Ref.~\cite{PERC}).

The results for the critical exponents using the 
quotients for the observables of Eq.~(\ref{OBS})  
from lattice pairs of sizes $(L,2L)$ measured where $Q_{\xi}=2$
are shown in table \ref{TEXPOISING}. We also report the values of 
the universal $g_4$ cumulant at the same points.
From this table, it is apparent that, with
the statistical error reached, an infinite volume extrapolation is
needed. This is especially clear for the $\eta$ exponent.

To perform this extrapolation, one should try to take into account the
corrections-to-scaling. For the Ising model, we shall show that the
leading order corrections are enough to obtain a fair extrapolation.

\begin{table}[t]
\begin{center}
\begin{tabular*}{\hsize}{@{\extracolsep{\fill}}rll}\hline\hline
\multicolumn{1}{c}{$L$}
&\multicolumn{1}{c}{$\beta_c(\xi/L,L)$}
&\multicolumn{1}{c}{$\beta_c(g_4,L)$}\\\hline\hline
8   &    0.2216246(32)  &   0.2218571(29)\\
12  &    0.2216438(13)  &   0.2217279(18)\\
16  &    0.2216487(9)   &   0.2216878(11)\\
24  &    0.2216519(5)   &   0.2216661(7) \\
32  &    0.22165314(45)   &   0.2216601(6) \\
48  &    0.22165438(23)   &   0.22165709(30) \\
64  &    0.22165432(19)   &   0.22165563(26) \\\hline\hline
\end{tabular*}
\caption{Crossing points of $\xi/L$ and $g_4$ for the Ising model 
obtained from lattice pairs $(L,2L)$.}
\label{CROSSISING}
\end{center}
\end{table}

We first need to evaluate the $\omega$ exponent. From the data of
table~\ref{TEXPOISING} it is difficult to obtain a sensible value of
$\omega$. We use the shift from the infinite volume critical coupling
of the crossing points of the scaling quantities $\xi/L$ and $g_4$
which are much more accurate. 
These points are shown in table~\ref{CROSSISING}.
Then we carry out a joint
fit of all values to the functional form given in Eq.~(\ref{BETACFIT}) (see
Refs.~\cite{OURFSS,ISDIL} for a more detailed exposition of the
method). Our criterium is to fit the data, using the full covariance matrix,
for lattices greater or equal than a given $L_\mathrm{min}$, increasing 
this minimum size until 
a stable value and a reasonable $\chi^2/\mathrm{d.o.f.}$ 
are found (see table~\ref{OMEGAJOINT}).  
We take as fitted parameters the corresponding to the first satisfactory
$L_\mathrm{min}$ with the statistical error of the fit discarding 
$L_\mathrm{min}$. 
In this case, we observe that $L_\mathrm{min}=16$ is
enough for our precision.  The value obtained $\omega=0.87(9)$ is
compatible with the computed using analytical
techniques~\cite{ZINN-JUSTIN}, or the recent experimental value 
$\omega=0.91(14)$~\cite{HENKEL}. Thus one can be confident that the
correction-to-scaling are mainly due to the leading term when $L\ge
16$. We should remark that for obtaining $\omega$ an estimation of $\nu$
must be used. 
For our accuracy,
a value of $\nu$ with an error at the
1\% level is enough. Therefore, we do not need a previous 
infinite volume extrapolation and
it is safe to take $\nu=0.63$ for this purpose.
The obtained infinite volume critical point, $\beta_{\mathrm c}$, 
is scarcely affected by the
uncertainty in $\omega$.

\begin{table}[t]
\begin{center}
\begin{tabular*}{\linewidth}{@{\extracolsep{\fill}}rcll}\hline\hline
\multicolumn{1}{c}{$L_\mathrm{min}$}
&$\chi^2/{\mathrm{d.o.f.}}$
&\multicolumn{1}{c}{$\omega$}        
&\multicolumn{1}{c}{$\beta_{\mathrm c}$}\\\hline
8  &10.50/10     &       0.934(14) &0.221654433(83)(18)\\
12 &10.06/8      &       0.938(24) &0.221654447(90)(7)\\
16 &1.796/6 & \underline{0.87}(4) &\underline{0.22165456}(11)(1)\\
24 &1.727/4 &  0.86(\underline 9)&0.22165459(\underline{15})(\underline{5})\\
\hline\hline
\end{tabular*}
\caption{Results of the infinite volume extrapolation of the crossing
points for $g_4$
and $\xi/L$, including data from $L\geq L_\mathrm{min}$ in the
Ising model. The first error bar in $\beta_\mathrm{c}$ is statistical
while the second is due to the uncertainty in $\omega$. We quote our 
preferred value and error bars by underlines. The extrapolation for 
$\beta_\mathrm c$ decreases when $\omega$ increases.}
\label{OMEGAJOINT}
\end{center}
\end{table}

\begin{table}[b!]
\begin{center}
\begin{tabular*}{\linewidth}{@{\extracolsep{\fill}}llllll}\hline\hline
Ref. & \multicolumn{1}{c}{$\beta_\mathrm{c}$} &\multicolumn{1}{c}{$\nu$} 
     & \multicolumn{1}{c}{$\eta$} & \multicolumn{1}{c}{$\beta$}
     & \multicolumn{1}{c}{$\gamma$}\\ \hline
\cite{LIVET} &0.2216544(10)     & \\
\cite{BLOTE} &0.2216546(10)     & 0.6301(8) & 0.037(3)\\
\cite{GUPTA} &0.221655(1)(1)    & 0.625(1)  & 0.025(6)\\
\cite{TALAPOV}&0.2216544(6)     & \\
\cite{HASENBUSCH}&              & 0.6299(3) & 0.0359(10)\\
This work  &0.22165456(15)(5) & 0.6294(5)(5)&0.0374(6)(6)
        &0.3265(3)(1)&1.2353(11)(14)\\
\hline\hline
\end{tabular*}
\caption{Previous Monte Carlo determination of critical 
quantities for the Ising model.}
\label{OTROSISING}
\end{center}
\end{table}

Using this estimate for $\omega$ we can perform a fit to the
Eq.~(\ref{QUO}). In the last row of table \ref{TEXPOISING} we present 
the extrapolation results for $\nu,\eta$ and $g_4$. 
In all cases we have used the criterium described above, in order to deal 
with higher order scaling corrections ($L_{\mathrm {min}}=12,16,16$
for $\nu$, $\eta$ and $g_4$, respectively). 

\begin{figure}[t!]
\begin{center}
\leavevmode
\epsfig{file=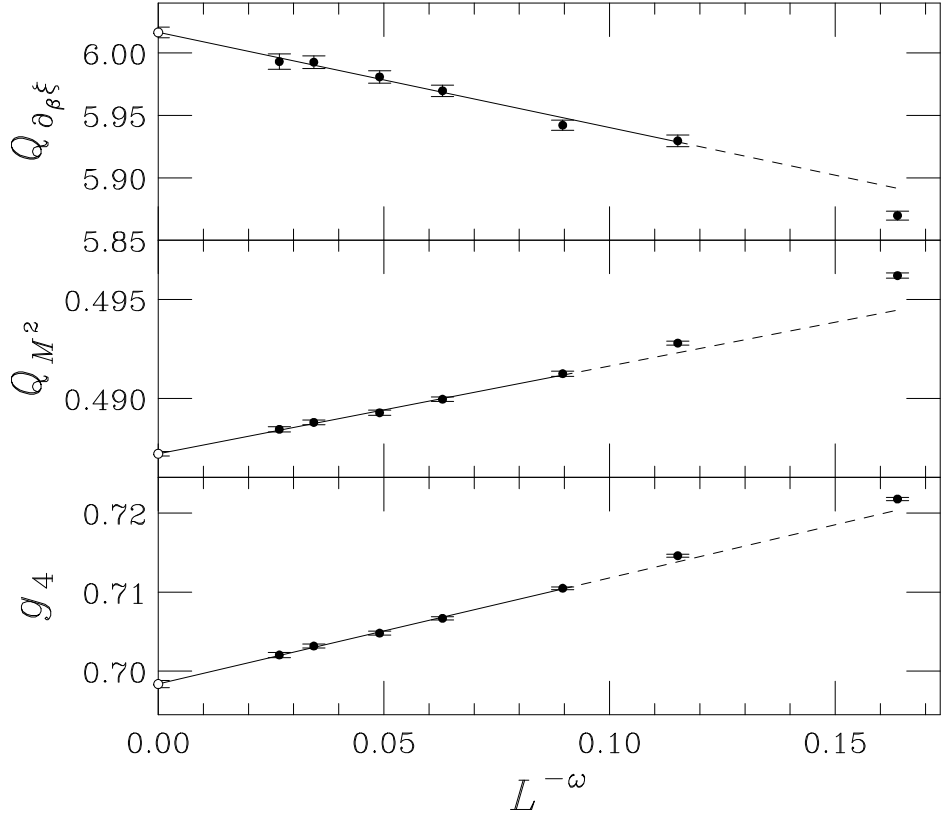,width=0.78\linewidth,angle=90}
\end{center}
\caption{Infinite volume extrapolation for $Q_{\partial_\beta \xi}=
2^{1+\frac{1}{\nu}}$, $Q_{M^2}=2^{-1-\eta}$, and $g_4$. The solid
lines correspond to fits with $\omega=0.87$ from $L \geq L_{\mathrm
{min}}$. A dashed line is plotted for $L<L_\mathrm{min}$.}
\label{EXPOISING}
\end{figure}

In Fig.~\ref{EXPOISING} we show graphically the fit quality for the
quotients used to obtain the exponents as well as for $g_4$ cumulant.

We remark that the pair $(L,2L)$ having a systematic
error in $\eta$ smaller than the final error in our extrapolation
has $2L=2000$ ($2L=800$ for $\nu$).
We recall that our largest pair has $2L=128$.

Values for the critical quantities obtained with MC  by other authors are
reported in table~\ref{OTROSISING}. 
For comparison, a recent series computation\cite{SALMAN}
yielded $\beta_{\mathrm{c}}=0.221659^{+2}_{-5}$.
Only in Ref.~\cite{BLOTE} an
infinite-volume extrapolation is considered for the critical
exponents. A different approach has been used in
Ref.~\cite{HASENBUSCH}, where the Hamiltonian is numerically tuned, in
order to make the ${\cal O} (L^{-\omega})$ corrections for cumulant
$g_4$ vanishing. This largely reduces the corrections-to-scaling for
the exponents.  However, the data of~\cite{HASENBUSCH} have been
analyzed as {\it if these coefficients, $A_O$, would be exactly
zero}. However, there still is an error associated to the uncertainty
in the {\it assumption} $A_{O}=0$, that has not been considered in
Ref.~\cite{HASENBUSCH}.  In fact, there is nothing special in the
value $A_{O}=0$, the only essential ingredient for the dramatic
reducing of the error estimate is to neglect the error in $A_O$.  Had
we disregarded the error in $A_{Q_O}$, we would obtain
$\nu=0.6294(2),\, \eta=0.0374(2)$.  However, we do not believe this to
be a valid procedure.

\section{Site percolation results}

The MC simulation in this case is rather different,
since one generates directly independent configurations. We
will work in the so-called canonical formulation in which the
probability of finding a hole in a given lattice site is
independent from the rest of sites.

It is very fast to generate the different configurations and most of computer
time is employed in tracing the clusters. We generate 32 millions of
samples for $L\leq 96$, 16 millions for $L=96,128$ and 4 millions for 
$L=192$.  As we need
individual measures for the $p$-extrapolation it is necessary to store
them on disk as they are obtained from different processors. 
In all cases we simulate at $p=0.3116$~\cite{STAUFFERBOOK}.

In table \ref{TEXPOPERC} we present the results for the exponents
$\nu$  and $\eta$  as well as
the $g_2$ and $g_4$ cumulants, obtained from different pairs of lattices.

\begin{table}[b]
\begin{center}
\begin{tabular*}{\hsize}{@{\extracolsep{\fill}}rllll}\hline\hline
\multicolumn{1}{c}{$L$} 
& \multicolumn{1}{c}{$\nu$}  
& \multicolumn{1}{c}{$\eta$} 
& \multicolumn{1}{c}{$g_2$} 
& \multicolumn{1}{c}{$g_4$} \\\hline
8   &    0.8802(6)    &   -0.01531(12) & 0.35395(11) &   0.72353(9)\\
12  &    0.8847(6)    &   -0.03230(12) & 0.35395(11) &   0.72353(9)\\
16  &    0.8825(7)    &   -0.03844(12) & 0.34854(10) &   0.72074(10)\\
24  &    0.8807(10)   &   -0.04267(12) & 0.34601(10) &   0.71695(8) \\
32  &    0.8809(10)   &   -0.04423(10) & 0.34559(10) &   0.71499(9) \\
48  &    0.8771(14)   &   -0.04531(12) & 0.34603(11) &   0.71290(9) \\
64  &    0.8757(17)   &   -0.04539(10) & 0.34638(10) &   0.71195(8) \\
96  &    0.8796(33)   &   -0.04554(20) & 0.34672(24) &   0.71124(20)\\\hline
$\infty$&0.8765(16)(2)&
-0.04602(27)(7)&0.34675(26)(6)&0.71052(21)(19)\\\hline
\hline
\end{tabular*}
\caption{Critical exponents for the  site percolation 
obtained from pairs of type $(L,2L)$ obtained at $Q_{\xi}=2$. We also
show the cumulants $g_4(L)$ and $g_2(L)$ at the same points. 
The last row correspond to an infinite volume extrapolation, showing
the statistical error (first bar) and that coming from the uncertainty
in $\omega$ (second bar).}
\label{TEXPOPERC}
\end{center}
\end{table}

\begin{table}[b!]
\begin{center}
\begin{tabular*}{\hsize}{@{\extracolsep{\fill}}rll}\hline\hline
\multicolumn{1}{c}{$L$}
&\multicolumn{1}{c}{$p_{\mathrm c}(\xi/L,L)$}
&\multicolumn{1}{c}{$p_{\mathrm c}(g_4,L)$}\\\hline
8     &  0.309761(7)   &    0.313201(11)  \\
12    &  0.311034(5)   &    0.312454(7) \\
16    &  0.3113614(36) &    0.3120770(49)\\
24    &  0.3115337(20) &    0.3117950(26)\\
32    &  0.3115788(14) &    0.3117007(19)\\
48    &  0.3115992(9)  &    0.3116390(12)\\
64    &  0.3116036(8)  &    0.3116214(12)\\
96    &  0.3116063(9)  &    0.3116122(11)\\\hline\hline
\end{tabular*}
\caption{Crossing points for $\xi/L$ and $g_4$ for the site percolation 
obtained from pairs of type $(L,2L)$.}
\label{CROSSPERC}
\end{center}
\end{table}

Regarding $p_\mathrm{c}$, we show in table~\ref{CROSSPERC} the
crossing points of $\xi/L$ and $g_4$ for pairs $(L,2L)$.  We find a
quite small, albeit significant, drift even in the largest lattices.
Notice that the values for the $\xi/L$ ($g_4$) crossing are monotonically
increasing (decreasing) with $L$. Thus, unless 
something weird happens with the scaling corrections, $p_{\mathrm c}$ is
bounded from above and below and one can readily extract
$p_{\mathrm c}=0.311609(3)$. However, a more precise $p_{\mathrm c}$ 
determination is possible using FSS techniques.
We proceed as before to make a joint fit to Eq.~(\ref{BETACFIT}) for all
the data in table~\ref{CROSSPERC}, excluding those for
$L<L_\mathrm{min}$. The results are presented in table~\ref{OMEGAJOINTPERC}.

\begin{table}[ht]
\begin{center}
\begin{tabular*}{\linewidth}{@{\extracolsep{\fill}}rcll}\hline\hline
$L_\mathrm{min}$     &$\chi^2/{\mathrm{d.o.f.}}$ 
&\multicolumn{1}{c}{$\omega$}
&\multicolumn{1}{c}{$p_{\mathrm {c}}$}\\\hline
24 &10.5/6 &   1.57(2)   &0.3116092(5)(2)\\
32 &0.65/4 &   \underline{1.62}(4)  &\underline{0.3116081}(7)(2)\\
48 &0.07/2 & 1.64(\underline{13})&0.3116075(\underline{11})(\underline{2})\\
\hline\hline
\end{tabular*}
\caption{Infinite volume extrapolation of the crossing
points for $g_4$
and $\xi/L$, including data from $L\geq L_\mathrm{min}$ in the
site percolation, in order to obtain $\omega$ and $p_{\mathrm {c}}$.}
\label{OMEGAJOINTPERC}
\end{center}
\end{table}

The value of $\omega=1.62(13)$ is remarkably larger than in most of the 3D
systems (slightly below 1), but agrees with the prediction of
$\epsilon$-expansions, and lies between the $\omega_\mathrm{2D}=2$ of
2D site percolation~\cite{NIENHUIS} and the four dimensional
value~\cite{PERC} $\omega_\mathrm{4D}=1.13(10)$. Moreover, our
value for $\omega$ is in agreement with other MC determinations
(see table~\ref{OTROSPERC}).

\begin{figure}[t!]
\begin{center}
\leavevmode
\epsfig{file=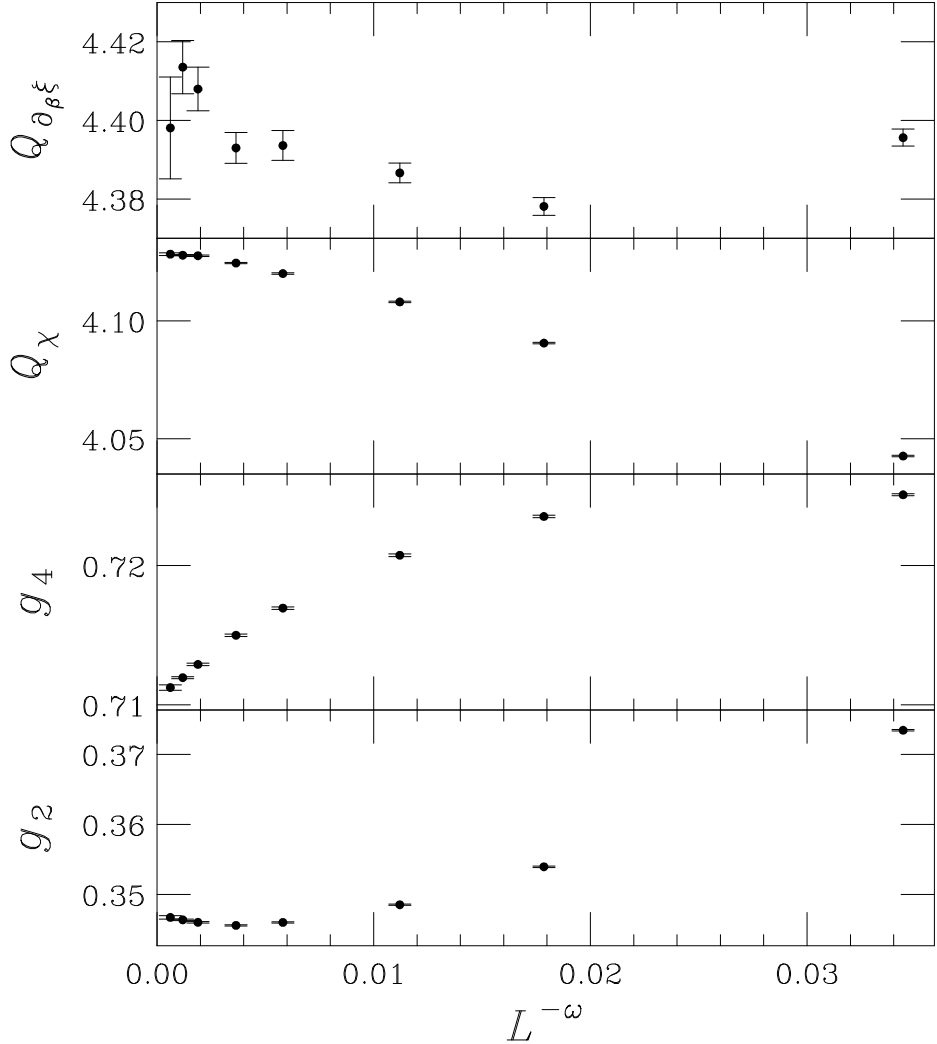,width=\linewidth,angle=90}
\end{center}
\caption{ $Q_{\partial_\beta \xi}=2^{1+\frac{1}{\nu}}$,
$Q_{\chi}=2^{2-\eta}$, $g_4$ and $g_2$ for pairs ($L,2L$) as  functions
of $L^{-\omega}$. In the plot we use $\omega=1.62$.}
\label{EXPOPERC}
\end{figure}

In Fig.~\ref{EXPOPERC} we display the quotients of $\chi$ and 
${\partial_\beta \xi}$, and the values of $g_4$ and $g_2$ as 
functions of $L^{-\omega}$ for $\omega=1.62$, measured at the points
where $Q_{\xi}=2$. The linear behaviour is much less clear than in the 
Ising case.
One should not be surprised by this fact since for such a large
$\omega$ it is very unlikely a clear separation between the leading
corrections-to-scaling (as $L^{-\omega}$) and the sub-leading ones. 
One should be specially worried with the analytical corrections that
for most operators go as $L^{-\gamma/\nu}\approx L^{-2}$.
A parameterization of the sub-leading corrections
is far from the present MC capacities.

Fortunately, $\omega$ is large enough to make the extrapolation almost
unnecessary. For $\nu$, we do not find significant deviations for
$L\ge 48$ and one could be tempted of simply averaging, obtaining
$\nu_\mathrm{mean}=0.8768(10)$. However, we find no reason to consider
as vanishing
the coefficient of $L^{-\omega}$, and this assumption underestimates
the errors. We find a non zero
value of $A_{Q_O}$ in the fits to Eq.~(\ref{QUO}) for
$\eta$ and the cumulants. In the last raw of table~\ref{TEXPOPERC}
we present the results of these fits as well as the
corresponding statistical errors, the second error bars corresponding
to the uncertainty in $\omega$. This $\omega$-error
allows to quantify the possible shift that could be 
expected if the dominant corrections-to-scaling were the analytical
ones, as the behaviour is basically linear with $\omega$. 
One simply has to add 2.5 times the $\omega$ induced error to the
central value for the extrapolation (the sign would be positive in the
four cases).  
For $\nu$, $\eta$ and $g_2$, one can conclude that the systematic
errors are hardly greater than the statistical one. For $g_4$ the
former could be twice the latter. Our final results can be contrasted
with other MC estimates in table \ref{OTROSPERC}.

\begin{table}[t]
\begin{center}
\begin{tabular*}{\linewidth}{@{\extracolsep{\fill}}lllll}\hline
Ref. & \multicolumn{1}{c}{$p_\mathrm{c}$} & \multicolumn{1}{c}{$\sigma$} 
     & \multicolumn{1}{c}{$\tau$}& \multicolumn{1}{c}{$\omega$}\\\hline\hline
\cite{GRASSBERGER}&0.311604(6)&         & 2.188(2) &        \\
\cite{LORENZ}&                &0.445(10) & 2.189(2) & 1.61(5)\\     
\cite{STAUFFER}&   0.311600(5)&         & 2.186(2) & 1.77(13)\\               
\cite{PERCOLA}&               &         &          & 1.4\\
\hline
This work &0.3116081(7)(2)  &0.4522(8)(1)&2.18906(6)(2)&1.62(13)\\
\hline    
\end{tabular*}
\end{center}
\caption{Previous Monte Carlo determination of critical quantities for the
3D site percolation.}
\label{OTROSPERC}
\end{table}

\section{Conclusions}

We have found that when measuring critical exponents and other
universal quantities with high precision (below the 0.1\%) with
finite-size scaling techniques, a proper consideration of the
corrections-to-scaling is mandatory.

We have studied two simple three dimensional models. The Ising model
shows corrections that can be parameterized with the leading
corrections-to-scaling term. It is possible to obtain a very safe 
infinite volume extrapolation that can be as far as 10 standard
deviations from the largest lattice's value.

In the site percolation, the behaviour is completely different. The
leading corrections-to-scaling cannot be easily isolated from the
higher order ones, since the first irrelevant exponent is very large.
However, its largeness makes the results on the largest lattices very
near the infinite volume limit, and the difficulties of the
extrapolation are not overwhelming.  We have also measured with high
precision the values of two universal cumulants ($g_4$, $g_2$). The
non-vanishing value of the latter shows that the susceptibility is
not a self-averaging quantity.

\section*{Acknowledgements}

We acknowledge interesting discussions with D. Stauffer.
We thank partial financial support from CICyT (AEN97-1708 and
AEN97-1693).  The computations have been carried out using the RTNN
machines at Universidad de Zaragoza and Universidad Complutense de
Madrid.

\end{document}